%% file: wyrowki-h2op-astroph.tex
\documentclass[traditabstract,letter,longauth]{aa} 
\usepackage{graphicx}
\usepackage{natbib}
\usepackage{verbatim}
\usepackage{txfonts}
\usepackage{ulem}
      \def\new#1 {{\bf #1 }}
      \def\cut#1 {\sout{#1} }

\def\cmsq  {$\hbox{{\rm cm}}^{-2}$}    

\def\OHP {OH$^+$}

\def\WAT {H$_2$O}
\def\WATET {H$_2^{18}$O}
\def\WATP {H$_2$O$^+$}

\def\hh{H$_2$}

\def\hho{H$_2$O}
\def\hhop{H$_2$O$^+$}
\def\hhhop{H$_3$O$^+$}

\include{definitions}

\begin{document}

\title{Variations in \WATP/\WAT\ ratios toward massive star-forming regions
       \thanks{{\it Herschel} is an ESA space observatory with science instruments 
       provided by European-led Principal Investigator consortia and with 
       important participation from NASA.}}


\author{
F.~Wyrowski \inst{ 1 }
\and  F.~van~der~Tak \inst{ 2 , 3 }
\and  F.~Herpin \inst{ 4 }
\and  A.~Baudry \inst{ 4 }
\and  S.~Bontemps \inst{ 4 }
\and  L.~Chavarria \inst{ 4 }
\and  W.~Frieswijk \inst{ 3 }
\and  T.~Jacq \inst{ 4 }
\and  M.~Marseille \inst{ 2 }
\and  R.~Shipman \inst{ 2 }
\and  E.F~van~Dishoeck \inst{ 5 , 6 }
\and  A.O.~Benz \inst{ 7 }
\and  P.~Caselli \inst{ 8 }
\and  M.R.~Hogerheijde \inst{ 6 }
\and  D.~Johnstone \inst{ 9 , 10 }
\and  R.~Liseau \inst{ 11 }
\and  R.~Bachiller \inst{ 12 }
\and  M.~Benedettini \inst{ 13 , 14 }
\and  E.~Bergin \inst{ 14 }
\and  P.~Bjerkeli \inst{ 11 }
\and  G.~Blake \inst{ 15 }
\and  J.~Braine \inst{ 4 }
\and  S.~Bruderer \inst{ 7 }
\and  J.~Cernicharo \inst{ 16 }
\and  C.~Codella \inst{ 13 }
\and  F.~Daniel \inst{ 17 , 18 }
\and  A.M.~di~Giorgio \inst{ 13 }
\and  C.~Dominik \inst{ 19 }
\and  S.D.~Doty \inst{ 20 }
\and  P.~Encrenaz \inst{ 21 }
\and  M.~Fich \inst{ 22 }
\and  A.~Fuente \inst{ 12 }
\and  T.~Giannini \inst{ 13 }
\and  J.R.~Goicoechea \inst{ 16 }
\and  Th.~de~Graauw \inst{ 23 }
\and  F.~Helmich \inst{ 2 }
\and  G.J.~Herczeg \inst{ 5 }
\and  J.K.~J{\o}rgensen \inst{ 24 }
\and  L.E.~Kristensen \inst{ 6 }
\and  B.~Larsson \inst{ 25 }
\and  D.~Lis \inst{ 15 }
\and  C.~McCoey \inst{ 22 }
\and  G.~Melnick \inst{ 26 }
\and  B.~Nisini \inst{ 13 }
\and  M.~Olberg \inst{ 11 }
\and  B.~Parise \inst{ 1 }
\and  J.C.~Pearson \inst{ 27 }
\and  R.~Plume \inst{ 28 }
\and  C.~Risacher \inst{ 2 }
\and  J.~Santiago \inst{ 12 }
\and  P.~Saraceno \inst{ 13 }
\and  M.~Tafalla \inst{ 12 }
\and  T.A.~van~Kempen \inst{ 26 }
\and  R.~Visser \inst{ 6 }
\and  S.~Wampfler \inst{ 7 }
\and  U.A.~Y{\i}ld{\i}z \inst{ 6 }
\and  J.H.~Black \inst{ 11 }
\and  E.~Falgarone \inst{ 17 }
\and  M.~Gerin \inst{ 17 }
\and  P.~Roelfsema \inst{ 2 }
\and  P.~Dieleman \inst{ 2 }
\and  D.~Beintema \inst{ 2 }
\and  A.~De~Jonge \inst{ 2 }
\and  N.~Whyborn \inst{ 23 }
\and  J.~Stutzki \inst{ 29 }
\and  V.~Ossenkopf \inst{ 29 }
}

\institute{
Max-Planck-Institut f\"{u}r Radioastronomie, Auf dem H\"{u}gel 69, 53121 Bonn, Germany
\and 
SRON Netherlands Institute for Space Research, PO Box 800, 9700 AV, Groningen, The Netherlands
\and 
Kapteyn Astronomical Institute, University of Groningen, PO Box 800, 9700 AV, Groningen, The Netherlands
\and 
Universit\'{e} de Bordeaux, Laboratoire d’Astrophysique de Bordeaux, France; CNRS/INSU, UMR 5804, Floirac, France
\and 
Max Planck Institut for Extraterestrische Physik, Garching, Germany
\and 
Leiden Observatory, Leiden University, PO Box 9513, 2300 RA Leiden, The Netherlands
\and 
Institute of Astronomy, ETH Zurich, 8093 Zurich, Switzerland
\and 
School of Physics and Astronomy, University of Leeds, Leeds LS2 9JT
\and 
National Research Council Canada, Herzberg Institute of Astrophysics, 5071 West Saanich Road, Victoria, BC V9E 2E7, Canada
\and 
Department of Physics and Astronomy, University of Victoria, Victoria, BC V8P 1A1, Canada
\and 
Department of Radio and Space Science, Chalmers University of Technology, Onsala Space Observatory, 439 92 Onsala, Sweden
\and 
IGN Observatorio Astron\'{o}mico Nacional, Apartado 1143, 28800 Alcal\'{a} de Henares, Spain
\and 
INAF - Istituto di Fisica dello Spazio Interplanetario, Area di Ricerca di Tor Vergata, via Fosso del Cavaliere 100, 00133 Roma, Italy
\and 
Department of Astronomy, The University of Michigan, 500 Church Street, Ann Arbor, MI 48109-1042, USA
\and 
California Institute of Technology, Division of Geological and Planetary Sciences, MS 150-21, Pasadena, CA 91125, USA
\and 
Department of Astrophysics, CAB, INTA-CSIC, Crta Torrej\'{o}n a Ajalvir km 4, 28850 Torrej\'{o}n de Ardoz, Spain
\and 
Observatoire de Paris-Meudon, LERMA UMR CNRS 8112, 5 place Jules Janssen, 92195 Meudon Cedex, France
\and 
Department of Molecular and Infrared Astrophysics, Consejo Superior de Investigaciones Cientificas, C/ Serrano 121, 28006 Madrid, Spain
\and 
Astronomical Institute Anton Pannekoek, University of Amsterdam, Kruislaan 403, 1098 SJ Amsterdam, The Netherlands 
\and 
Department of Physics and Astronomy, Denison University, Granville, OH, 43023, USA
\and 
LERMA and UMR 8112 du CNRS, Observatoire de Paris, 61 Av. de l'Observatoire, 75014 Paris, France
\and 
University of Waterloo, Department of Physics and Astronomy, Waterloo, Ontario, Canada
\and 
Atacama Large Millimeter/Submillimeter Array, Joint ALMA Office, Santiago, Chile 
\and 
Centre for Star and Planet Formation, Natural History Museum of Denmark, University of Copenhagen, {\O}ster Voldgade 5-7, DK-1350 Copenhagen, Denmark
\and 
Department of Astronomy, Stockholm University, AlbaNova, 106 91 Stockholm, Sweden
\and 
Harvard-Smithsonian Center for Astrophysics, 60 Garden Street, MS 42, Cambridge, MA 02138, USA
\and 
Jet Propulsion Laboratory, California Institute of Technology, Pasadena, CA 91109, USA
\and 
Department of Physics and Astronomy, University of Calgary, Calgary, T2N 1N4, AB, Canada
\and 
Physikalisches Institut, Universit\"{a}t zu K\"{o}ln, Z\"{u}lpicher Str. 77, 50937 K\"{o}ln, Germany
}


\date{Received / Accepted}


\abstract {
  Early results from the {\it Herschel} Space Observatory revealed the water
  cation \WATP\ to be an abundant ingredient of the interstellar medium.
  Here we present new observations of the \hho\ and \hhop\ lines at
  1113.3 and 1115.2\,GHz using the {\it Herschel} Space Observatory
  toward a sample of high-mass star-forming regions to observationally
  study the relation between \WAT\ and \WATP. Nine out of ten sources
  show absorption from \WATP\ in a range of environments: the
  molecular clumps surrounding the forming and newly formed massive
  stars, bright high-velocity outflows associated with the massive
  protostars, and unrelated low-density clouds along the line of
  sight. Column densities per velocity component of \WATP\ are found
  in the range of $10^{12}$ to a few $10^{13}$~\cmsq.  The highest
  $N$(\WATP) column densities are found in the outflows of the
  sources. The ratios of \WATP/\WAT\ are determined in a range from 0.01 to a
  few and are found to differ strongly between the observed
  environments with much lower ratios in the massive (proto)cluster
  envelopes (0.01-0.1) than in outflows and diffuse
  clouds. Remarkably, even for source components detected in \WAT\ in
  emission, \WATP\ is still seen in absorption.
}

\keywords{ISM: molecules -- 
          ISM: clouds --
          Submillimeter: ISM --
          Stars: formation
                 }

\authorrunning{Wyrowski et al.}

\maketitle

\section{\label{intro}Introduction}

One of the unique ESA {\it Herschel} Space Observatory (Pilbratt et
al. 2010) science fields is the observation of thermal lines of
interstellar water and other hydrides.
Hydrides have small reduced masses, so their rotational lines lie
at short submillimeter wavelengths, which are almost unobservable from
the ground \citep{phillips+vastel2003}.
Early results from the first months of observations show the
scientific potential of these studies \citep[e.g.][for water in a
massive star-forming region]{vdtak+2010}. Interestingly, these early
results also revealed the water cation \WATP, which was seen by
{\it Herschel} for the first time, as an abundant
ingredient of the interstellar medium
\citep{ossenkopf+2010,gerin}. The ortho ground-state line of \WATP\
was even detected in external galaxies and found to be stronger than
the para ground-state water line \citep{weiss,vdwerf}. These early
results indicate that \WATP\ originates mainly from low-density gas of
diffuse interstellar clouds.

Within the ``Water in star-forming regions with {\it Herschel}
(WISH)'' \citep{vandishoeck} {\it Herschel} key program, a sample of
about 20 massive star-forming regions (SFR) covering a wide range of
evolutionary stages is observed in a variety of water lines. One of
the \WATP\ ortho ground-state doublet lines lies close in frequency to
the \WAT\ para ground-state line and is observed as well. This allows
us to present here a detailed comparison of water and ionized water
column densities in a larger sample of sources to study relative
abundance variations of \WATP\ and \WAT\ in different interstellar
environments.

\section{\label{obs}Observations and data reduction}

The sources were observed with the Heterodyne Instrument for the
Far-Infrared (HIFI, De Graauw et al. 2010) onboard the {\it Herschel} Space
Observatory (Pilbratt et al. 2010) on 2010 March 3-5 and April 17.
Double-beam switch observations (throw of 2.5 arcminutes) have
been performed in the double sideband mode using the 4b receiver
band. The pointing coordinates of the observed sample are given in
Table~\ref{table:sources}.
Data were taken in two polarizations with the
acousto-optical wide-band spectrometer (WBS), which covers 4--8~GHz in
four sub-bands, each approximately 1.1 GHz wide. Its Nyquist
resolution is approximately 1.1~MHz (0.30 km s$^{-1}$). 
Four species have been observed simultaneously with the WBS: 
  p-H$_2$O (1113.3 GHz, USB) and p-H$_2$$^{18}$O $1_{11} - 0_{00}$
  (1101.7 GHz, LSB), $^{13}$CO 10-9 (1101.3 GHz, LSB), and
  o-H$_2$O$^+$ $1_{11}-0_{00}$ (1115.2 GHz for the strongest HF
  component, USB).

The system temperatures for our data were around 350~K. Integration
time was 601s. Calibration of the raw data onto $T_{\mathrm A}$ scale
was performed by the in-orbit system (Roelfsema et al 2010);
the conversion to $T_{\mathrm{MB}}$ was done with a beam efficiency of
0.7. The {\it Herschel} full-beam-at-half-maximum at this frequency was
assumed to be the theoretical one (20"). Currently, the flux scale is
accurate to 5\%. An rms of 90 mK has been reached.

Data calibration was performed in the {\it Herschel} Interactive
Processing Environment (HIPE) version 2.8. The velocity uncertainty in
the current version of the pipeline is up to 2 \kms, depending on
target direction and observation epoch. Further analysis was done
within the CLASS package. After inspection, the data from the two
polarizations were averaged together. The continuum level in the data
was divided by 2, because the original calibration was done for the
line emission originating from only one receiver sideband.

 \begin{table}
  \caption{\label{table:sources} Sources observed in the 1110~GHz setup. }
  {\begin{tabular}{lllcccc}
  \hline\hline
  Source        & Ra (J2000)    & Dec       & V$_{\rm lsr}$ &  L$_{\rm bol}$  \\
                & ( h m s ) & ( $^\circ$ $^\prime$ $^{\prime\prime}$ )      & (km s$^{-1}$)  &  ($L_\odot$/$10^4$)    \\
  \hline
AFGL2591            & 20 29 24.7    &   +40 11 19   & $-$5.5   & 2.0     \\
DR21(OH)            & 20 39 00.8    &   +42 22 48   & $-$4.5  & 1.7      \\
G29.96$-$0.02       & 18 46 03.8    & $-$02 39 22   & +98.7   & 12       \\
G31.41+0.31         & 18 47 34.3    & $-$01 12 46   & +98.8   & 18        \\
G34.26+0.15         & 18 53 18.6    &   +01 14 58   & +57.2   &  28       \\
NGC7538-IRS1        & 23 13 45.3    &   +61 28 10   & $-$57.4  & 20     \\
W3-IRS5             & 02 25 40.6    &   +62 05 51   & $-$38.4  & 17     \\
W33A                & 18 14 39.1    & $-$17 52 07   & +37.5 & 1.0         \\
W43-MM1             & 18 47 47.0    & $-$01 54 28   & +98.8  & 2.3        \\
W51N E1             & 19 23 40.0    &   +14 30 51   & +59.0   &  10-100  \\
\hline
\end{tabular}}
\end{table}

\section{Results}
\label{sec:results}

\begin{figure}[ht]
\begin{center}
\includegraphics[height=0.48\textwidth,angle=-90]{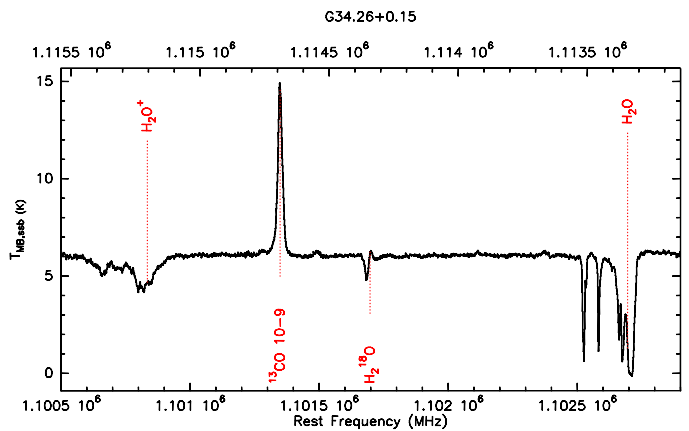}
\caption{\label{fig:g34} Example DSB spectrum at 1110~GHz towards
  G34.26 showing all lines covered in this spectral setup at the
  $V_{\mathrm LSR}$ of the source. {The redshifted \WAT\ and
    \WATP\ features are caused by the line-of-sight absorption components.}
  The lower and upper scales give the LSB and USB frequency scales,
  respectively. }
\end{center}
\end{figure}

In Fig.~\ref{fig:g34} the DSB WBS spectrum towards G34.26+0.15 is shown as
a typical example for the data that are analyzed in this study.  \WATP\
is detected in all sources except NGC7538~IRS1 and in all cases seen in
absorption, similar to the previous detections
\citep{ossenkopf+2010,gerin}. \WAT\ and \WATET\ on the other hand show
in many cases both absorption and emission line components. Many
sources show several velocity components owing to absorption
from diffuse clouds from different spiral arms on the lines of sight (LOS),
which complicates the interpretation of the \WATP\ spectra because of its
complex hyperfine structure \citep[HFS, see][for
details]{strahan+1986,muertz+1998}. Several sources
show saturated \WAT\ absorption down to the 0~K level 
(e.g. Figs.~\ref{fig:g34}-\ref{fig:w51}), demonstrating that the
continuum level in the spectra is measured reliably and that the
sideband gain ratio is 1. This is especially important because most
of the analysis is based on the detected absorption features for which
the line-to-continuum ratio is the relevant observing parameter. Most of the
sources show broad wing emission ($\Delta V>10$~\kms) as an indication of powerful
outflows, in \WAT\ mostly seen in emission, while in \WATP\ outflows
are detected as broad (due to the blended HFS)
blueshifted absorption features in front of the strong dust continuum
emission.

\section{Analysis}
\label{sec:analysis}

Because all lines for a given source are observed simultaneously in the
DSB spectra, they will only have a small relative error in their
intensities and velocities independent of the calibration. 
%
%
Here, the rest frequencies from \citet{muertz+1998} are used, which have a quoted accuracy of
2~MHz. A comparison of the \WAT\ and \WATP\ velocities from LOS
absorption features shows a small scatter of $\pm$2~\kms, hence the
resulting accuracy of the measured \WATP\ velocities is sufficient to
associate them with known velocity components of the observed
sources.

\begin{figure}[b]
\begin{center}
\includegraphics[height=0.48\textwidth,angle=-90]{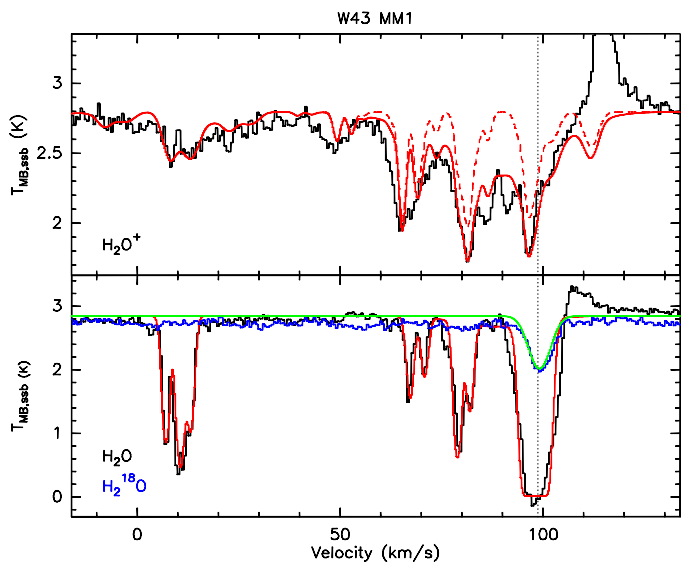}
\caption{\label{fig:w43} XCLASS fits to the
  water (lower panel) and water ion line profiles (upper panel) in W43 MM1 shown in red solid lines. The fit to the \WATET\ line (blue spectrum) is shown in green. An \WATP\ fit without 
  a broad outflow component is shown with red dots. The systemic velocity
  of the source is indicated by a dotted line. The strong emission
  line in the \WATP\ spectrum is $^{13}$CO (10-9) from the other sideband.}
\end{center}
\end{figure}

\begin{figure}[t]
\begin{center}
\includegraphics[height=0.48\textwidth,angle=-90]{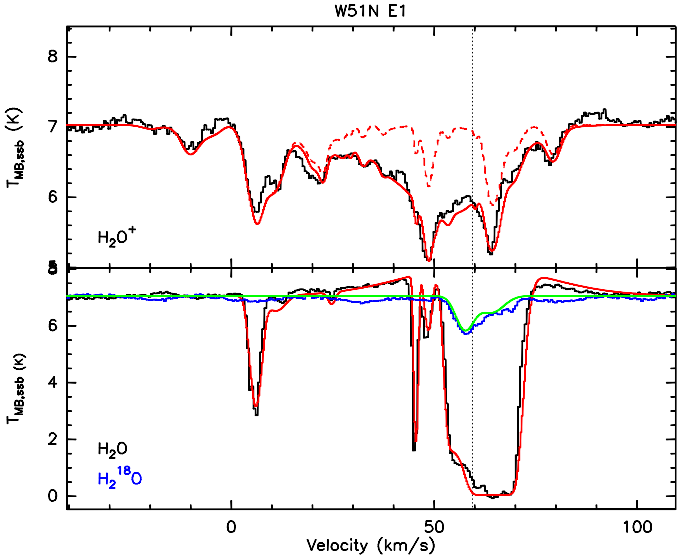}
\caption{\label{fig:w51} Same as Fig.~\ref{fig:w43} for W51N E1.}
\end{center}
\end{figure}

To separate the various and often blended velocity components and the \WATP\ HFS, we used the XCLASS fitting tool
\citep[][and references therein]{comito+2005}, which allows us to obtain
multi-component LTE fits of emission and
absorption components and which takes the HFS -- which extends over 40~\kms\ \citep{muertz+1998} -- into account. The input parameters for the fits are the
excitation temperature, column density, source size, source velocity
and velocity width.  
For the absorption components, a background
brightness temperature of 15-25 K was derived from the measured
continuum temperatures assuming source sizes as given in
Table~\ref{t:fits}.

The \WAT\ absorption components with high velocity offsets, which are
likely diffuse LOS clouds, were fitted with an excitation temperature
of 2.7~K. However, the diffuse Galactic background radiation might
increase the excitation temperatures of the submillimeter lines to
values of about 5 K.  For absorption components that are likely
associated with the massive SFRs we used a fixed value of 5~K, which
is clearly below the background temperature, to get an absorption in
the fit. However, the corresponding column densities for temperatures
below 10~K depend only weakly on the assumed $T_{\mathrm EX}$.  For
the emission components we used a fixed value of 50~K. The source size
is assumed to be much bigger than the beam.  Only for the 57.5~\kms\
component toward W51N E1 a size of 35\arcsec\ is assumed. Judging
from its strong \WATET\ absorption, it has a very high \WAT\ optical
depth, but still does not absorb the continuum down to 0~K and
therefore requires a filing factor smaller than one.  For the column
density calculations we used an ortho-to-para ratio of 3:1 for
\WATP.

The fit results obtained for \WAT\ were used as the starting point to
fit the \WATP\ spectra, which are more complex owing to the \WATP\ HFS.
because no emission is seen in \WATP, we changed any component seen in
emission in \WAT\ into an absorption component for \WATP\ by lowering
its excitation temperature to 5~K.  The physical parameters for each
component were then fine-tuned to fit the observed \WATP\ spectra. In
cases where the corresponding component was not detected in \WATP, the
highest column density consistent with a non-detection was chosen to
derive an upper limit. In a few cases, the observed \WAT\ components
were not sufficient to account for all the \WATP\
absorption. In Fig.~\ref{fig:w43} e.g. there is no indication of an
\WAT\ outflow component and the blue-shifted LOS absorptions are quite
narrow, while the \WATP\ absorption is very broad, even considering its
HFS, so that an additional blueshifted, broad ``outflow'' component
($\Delta V> 10$~\kms) was added to reproduce the observed \WATP\
spectrum.

Two examples for the resulting fits are shown in Figs.~\ref{fig:w43}
and \ref{fig:w51}. The corresponding fit parameters for these sources
are given in Table~\ref{t:fits}.  The range in $N$(\WATP) of $10^{12}$
to a few $10^{13}$~\cmsq\ is much smaller than the range in $N$(\WAT)
of a few $10^{12}$ to several $10^{14}$~\cmsq.
Similar results are obtained for other sources in our sample and will be discussed elsewhere and in the following section.

\begin{table}[t]
\caption{\WAT, \WATET\ and \WATP\ fit results of velocity components
         associated with the massive star-forming clumps W43 and W51
         shown in Figs.~\ref{fig:w43} and \ref{fig:w51}.
}
\label{t:fits} 
\begin{center}
 \begin{tabular}{lrrrr}
 \hline\hline 
 Mol. / Source & $T_{\mathrm ex}$ & $N/10^{12}$& $\Delta V$ & $V_{\mathrm{lsr}}$ \\ 
 & (K)  & (\cmsq) & (km~s$^{-1}$) & (km~s$^{-1}$) \\
\hline 
\input{molfit-2sources-nosize.tex}
\hline
\end{tabular}
\end{center}
\end{table}

\section{Discussion and  conclusions}

\begin{figure}[ht]
\begin{center}
\includegraphics[height=0.48\textwidth,angle=-90]{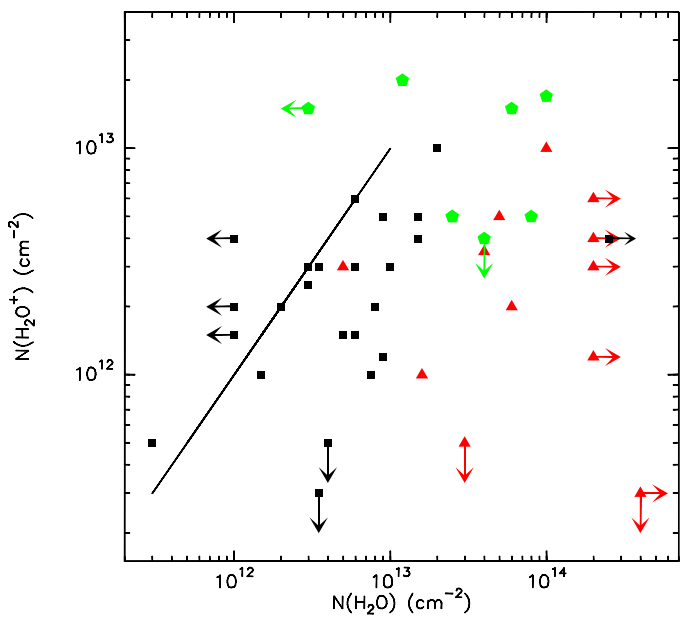}
\caption{\label{fig:corr} Comparison of \WAT\ and \WATP\ column
  densities of different line components in the spectra: the black
  squares give diffuse lines of sight, red triangles the envelopes of
  the massive star-forming clumps, and green filled dots the outflow
  components ($\Delta V> 10$~\kms). As a reference, the solid line
  shows N(\WATP)=N(\WAT). Some of the low (high) column densities are
  upper (lower) limits and are indicated by arrows.}
\end{center}
\end{figure}

An overview of the fit results is shown in Fig.~\ref{fig:corr},  in
which we plot for each velocity component the corresponding \WAT\ and \WATP\
column densities. Some of the lower column densities are
upper limits while some of the high \WAT\ column densities are lower
limits because of the saturation in the \WAT\ absorption lines. In these
cases the true column density, estimated from the observed \WATET\
lines and assuming a $^{16}$O/$^{18}$O ratio in the range from
250--560 \citep{wilson+rood1994} will be higher by up to a factor
5. Figure~\ref{fig:corr} shows that compared to the properties of
diffuse clouds \citep[see also][]{gerin,ossenkopf+2010,weiss}, in which
$N$(\WATP) is closer to $N$(\WAT), significantly higher
$N$(\WAT)/$N$(\WATP) is found for velocity components likely
originating from the envelopes and bright high-velocity outflows of
the massive star-forming clumps.  Even in the outflows and envelopes of
the massive star-forming regions there is still a variation of
$N$(\WATP). 
The highest $N$(\WATP) column densities are found in
the outflows.

After the first indication that \WATP\ is present in the outflow of DR21
\citep{ossenkopf+2010}, it is clearly detected here in several
outflows. Some outflows are even more prominent in
\WATP\ than in \WAT\ (e.g. outflow components above the diffuse cloud
region of Fig.~\ref{fig:corr}), although in some cases it is
difficult to disentangle the blueshifted outflow and spiral arm
absorption.

In Fig.~\ref{fig:corr} one cloud treated in the literature as
diffuse LOS cloud toward W51 \citep[64~\kms\
component,][]{sollins+2004} has a high \WAT/\WATP\ ratio of 50,
which indicates that it might rather be associated with the W51 clump
envelope.  Strong ortho-water absorption of this component was already
seen with SWAS by \citet{neufeld+2002}.  Hence the \WAT/\WATP\ ratio
might be useful for the classification of absorption components.

A key result from our observation of a quite significant sample is
that \WATP\ is always seen in absorption, even when originating from
outflows or envelopes, in which water is seen in several cases in
emission. This finding might point to the origin of \WATP\ from
low-density environments, but alternatively, because \WATP\ is a highly
reactive molecular ion that reacts rapidly with \hh\ and electrons,
inelastic collisions are very ineffective at exciting it into
emission, regardless of the density.

The new detections of \WATP\ toward Galactic star-forming
regions presented here together with the recent detection of \OHP\
from the ground \citep{wyrowski} and from space
\citep{gerin,benz,bruderer} are an important confirmation of the
gas-phase route to water. The \WATP\ lines are stronger than the
\hhhop\ lines in the same sources (some of is caused by
differences in spectroscopic properties), which is surprising, because \hhop\ is
expected to react fast with \hh\ into \hhhop, which recombines
with electrons to produce \WAT. This puzzle is even more
pronounced in recent {\it Herschel}/HIFI observations in diffuse clouds
\citep{gerin} and active galactic nuclei \citep{weiss}, where \hhop\ is
even more abundant than \hho\ itself. While for AGN strong X-ray
and/or UV radiation is likely to dominate the chemistry
\citep{vdwerf}, environments without radiation sources require other
solutions. One solution might be that in the outer envelopes of the
massive SFRs \WAT\ is freezing out onto grains
whereas --in the case of positively charged grains -- \WATP\ is much less affected
by freeze-out.

The main destruction routes of \WATP\ are dissociative recombination
(into OH) and reaction with \hh\ (into \hhhop\ and \WAT).  The high \WATP/\WAT\
ratio observed in the diffuse components implies that the first channel
is faster than the second.  In gas where all hydrogen is in molecular
form, the electron fraction is $10^{-4}$ at most when all carbon is
ionized, which is not enough to make recombination faster than the
reaction with \hh.  Our observations
therefore imply that a significant fraction of the hydrogen in the
outflows is in atomic form.  The same conclusion applies to diffuse
clouds, where UV radiation causes partial dissociation of \hh\
\citep{gerin}, and also to AGN, where X-rays are responsible
\citep{vdwerf}.  For molecular outflows the most likely
mechanism to dissociate \hh\ is by fast (J-type) shocks.  The required
shock velocities of 30--40~\kms\ are easily reached in the powerful
outflows of the sources.

Models of dense PDRs \citep[e.g.][]{sternberg+dalgarno1995} predict
$\rm{OH^+ + H_2 \rightarrow H_2O^+}$. In the outflow-walls scenario
  \citep{bruderer+2009} this leads to a thin PDR layer along the
  outflow wall, where FUV heats and ionizes the gas. New modeling of
  Bruderer et al. (2010, subm.) of hydrides including \WATP\ predicts
  the abundance of \WATP\ to be enhanced by four orders of magnitude
  along the outflow compared to the envelope, which then could explain
  the high column densities of \WATP\ in the outflow components.

\bibliographystyle{aa}
\bibliography{hifi-h2op}

\acknowledgements{The XCLASS program
  (http://www.astro.uni-koeln.de/projects/schilke/XCLASS) was used,
  which accesses the CDMS (http://www.cdms.de) and JPL
  (http://spec.jpl.nasa.gov) molecular data bases.
HIFI has been designed and built by a consortium of
institutes and university departments from across Europe, Canada and the
United States under the leadership of SRON Netherlands Institute for Space
Research, Groningen, The Netherlands and with major contributions from
Germany, France and the US. Consortium members are: Canada: CSA,
U.Waterloo; France: CESR, LAB, LERMA, IRAM; Germany: KOSMA,
MPIfR, MPS; Ireland, NUI Maynooth; Italy: ASI, IFSI-INAF, Osservatorio
Astrofisico di Arcetri- INAF; Netherlands: SRON, TUD; Poland: CAMK, CBK;
Spain: Observatorio Astron´omico Nacional (IGN), Centro de Astrobiologia
(CSIC-INTA). Sweden: Chalmers University of Technology - MC2, RSS \&
GARD; Onsala Space Observatory; Swedish National Space Board, Stockholm
University - Stockholm Observatory; Switzerland: ETH Zurich, FHNW; USA:
Caltech, JPL, NHSC.
HCSS / HSpot / HIPE is a joint development (are joint developments) by
the {\it Herschel} Science Ground Segment Consortium, consisting of ESA, the
NASA {\it Herschel} Science Center, and the HIFI, PACS and SPIRE consortia.
We would like to thank the referee, Christian Henkel, for a thorough
review of the manuscript and helpful comments. 
}

\end{document}

%% file: definitions.tex
\def\folio{\ifnum\pageno=1\nopagenumbers\else\number\pageno\fi}

\def\lax    {\ifmmode{_<\atop^{\sim}}\else{${_<\atop^{\sim}}$}\fi}
\def\gax    {\ifmmode{_>\atop^{\sim}}\else{${_>\atop^{\sim}}$}\fi}
\newbox\grsign      \setbox\grsign=\hbox{$>$} 
\newdimen\grdimen   \grdimen=\ht\grsign
\newbox\simgreatbox \setbox\simgreatbox=\hbox{\raise.5ex\hbox{$>$}\llap
                        {\lower.5ex\hbox{$\sim$}}}\ht1=\grdimen\dp1=0pt
\newbox\simlessbox  \setbox\simlessbox =\hbox{\raise.5ex\hbox{$<$}\llap
                        {\lower.5ex\hbox{$\sim$}}}\ht2=\grdimen\dp2=0pt

\newbox\grsign \setbox\grsign=\hbox{$>$} \newdimen\grdimen \grdimen=\ht\grsign
\newbox\laxbox \newbox\gaxbox
\setbox\gaxbox=\hbox{\raise.5ex\hbox{$>$}\llap
     {\lower.5ex\hbox{$\sim$}}}\ht1=\grdimen\dp1=0pt
\setbox\laxbox=\hbox{\raise.5ex\hbox{$<$}\llap
     {\lower.5ex\hbox{$\sim$}}}\ht2=\grdimen\dp2=0pt
\def\gax{\mathrel{\copy\gaxbox}}
\def\lax{\mathrel{\copy\laxbox}}
\def\boxit#1    {\vbox{\hrule\hbox{\vrule\kern3pt
                  \vbox{\kern3pt#1\kern3pt}\kern3pt\vrule}\hrule}}
\def\h      {\ifmmode{^{\rm h}}\else{$^{\rm h}$}\fi}
\def\m      {\ifmmode{^{\rm m}}\else{$^{\rm m}$}\fi}
\def\s      {\ifmmode{^{\rm s}}\else{$^{\rm s}$}\fi}
\def\decas    {\ifmmode{{\rlap.}{''}}\else{${\rlap.}{''}$}\fi}
\def\mum     {\ifmmode{\mu{\rm m}}\else{$\mu{\rm m}$}\fi}
\def\s      {\ifmmode{^{\rm s}}\else{$^{\rm s}$}\fi}
\def\deg      {\ifmmode{^{\circ}}\else{$^{\circ}$}\fi}
\def\as     {\ifmmode {\rlap.}$\,$''$\,$\! \else ${\rlap.}$\,$''$\,$\!$\fi}
\def\decsec  {\ifmmode {\rlap.}$\,$^{s}$\,$\! \else ${\rlap.}$\,$^{s}$\,$\!$\fi}\def\decs  {\ifmmode {\rlap.}$\,$^{s}$\,$\! \else ${\rlap.}$\,$^{s}$\,$\!$\fi}

\def\kms    {\ifmmode{{\rm km~s}^{-1}}\else{km~s$^{-1}$}\fi}

\def\Mspy   {\ifmmode {M_{\odot} {\rm yr}^{-1}} \else $M_{\odot}$~yr$^{-1}$\fi}
\def\Mdot   {\ifmmode {\dot M} \else $\dot M$\fi}
\def\mhd    {\ifmmode {n_{{\rm H}_2}} \else $n_{{\rm H}_2}$\fi}
\def\mhcd   {\ifmmode {N_{{\rm H}_2}} \else $N_{{\rm H}_2}$\fi}

\def\El      {\ifmmode{E_{\ell}}\else{$E_{\ell}$}\fi}
\def\beam    {\ifmmode{\theta_{\rm B}}\else{$\theta_{\rm B}$}\fi}
\def\mjyb   {\ifmmode {{\rm mJy~beam}^{-1}} \else{mJy~beam$^{-1}$}\fi}
\def\mujyb   {\ifmmode {\mu{\rm Jy~beam}^{-1}} \else{$\mu$Jy~beam$^{-1}$}\fi}
\def\Trot   {\ifmmode{T_{\rm rot}}\else$T_{\rm rot}$\fi}    
    
\def\Teff   {\ifmmode{T_{\rm eff}}\else$T_{\rm eff}$\fi}

\def\ITRS   {\ifmmode{\smallint {\rm T}_{R}^{*}dv}\else{$\smallint 
{\rm T}_{R}^{*}dv$}\fi}
\def\ITRS   {\ifmmode{\smallint {\rm T}_{R}^{*}dv}\else{$\smallint 
{\rm T}_{R}^{*}dv$}\fi}
\def\ITAS   {\ifmmode{\smallint {\rm T}_{A}^{*}dv}\else{$\smallint 
{\rm T}_{A}^{*}dv$}\fi}

\def\hh         {H$_2$}

\def\hho        {H$_2$O}

\def\lefttitle#1  {\noindent \hangindent=18.0pt \hangafter=1 {#1} \par}
\def\vol#1  {{\bf {#1}{\rm,}\ }}
\font\tenssb=cmssbx10
\font\tenbf=cmbx10
\font\sevenbf=cmbx8
\font\fivebf=cmbx6
\def\unetdemi    {\smallskipamount=6pt plus2pt minus2pt
                  \medskipamount=12pt plus4pt minus4pt
                  \bigskipamount=24pt plus8pt minus8pt
                  \normalbaselineskip=16pt plus0pt minus0pt
                  \normallineskip=2pt
                  \normallineskiplimit=0pt
                  \jot=6pt
                  {\def\smallskip {\vskip\smallskipamount}}
                  {\def\medskip   {\vskip\medskipamount}}
                  {\def\bigskip   {\vskip\bigskipamount}}
                  {\setbox\strutbox=\hbox{\vrule 
                    height17.0pt depth7.0pt width 0pt}}
                  \parskip 12.0pt
                  \normalbaselines}
\def\smallerspace {\smallskipamount=3pt plus0pt minus0pt
                  \medskipamount=6pt plus0pt minus0pt
                  \bigskipamount=10.5pt plus0pt minus0pt
                  \normalbaselineskip=10.5pt plus0pt minus0pt
                  \normallineskip=1pt
                  \normallineskiplimit=0pt
                  \jot=3pt
                  {\def\smallskip {\vskip\smallskipamount}}
                  {\def\medskip   {\vskip\medskipamount}}
                  {\def\bigskip   {\vskip\bigskipamount}}
                  {\setbox\strutbox=\hbox{\vrule 
                    height8.5pt depth3.5pt width 0pt}}
                  \parskip 0pt
                  \normalbaselines}
\def\memospace    {\smallskipamount=4pt plus1pt minus1pt
                  \medskipamount=6pt plus2pt minus2pt
                  \bigskipamount=14pt plus6pt minus6pt
                  \normalbaselineskip=14pt plus0pt minus0pt
                  \normallineskip=1pt
                  \normallineskiplimit=0pt
                  \jot=4pt
                  {\def\smallskip {\vskip\smallskipamount}}
                  {\def\medskip   {\vskip\medskipamount}}
                  {\def\bigskip   {\vskip\bigskipamount}}
                  {\setbox\strutbox=\hbox{\vrule 
                    height17.0pt depth7.0pt width 0pt}}
                  \parskip 2.0pt
                  \normalbaselines}
\def\memowidespace    {\smallskipamount=5pt plus1pt minus1pt
                  \medskipamount=7.5pt plus2pt minus2pt
                  \bigskipamount=17.5pt plus6pt minus6pt
                  \normalbaselineskip=17.0pt plus0pt minus0pt
                  \normallineskip=1.25pt
                  \normallineskiplimit=0pt
                  \jot=5pt
                  {\def\smallskip {\vskip\smallskipamount}}
                  {\def\medskip   {\vskip\medskipamount}}
                  {\def\bigskip   {\vskip\bigskipamount}}
                  {\setbox\strutbox=\hbox{\vrule 
                    height21.25pt depth8.75pt width 0pt}}
                  \parskip 2.5pt
                  \normalbaselines}

%% file: molfit-2sources-nosize.tex
   \WAT\       / W43 MM1&  2.7 &   6.0 &  2.0 &   7.0 \\ 
                        &  2.7 &   9.0 &  2.0 &  10.5 \\ 
                        &  2.7 &   5.0 &  2.0 &  13.0 \\ 
                        &  2.7 &   3.0 &  2.0 &  66.5 \\ 
                        &  2.7 &   2.0 &  2.0 &  70.0 \\ 
                        &  2.7 &   7.5 &  2.0 &  78.0 \\ 
                        &  2.7 &   3.5 &  2.0 &  81.0 \\ 
                        &  5.0 &   3.0 & 20.0 &  87.0 \\ 
                        &  5.0 & 200.0 &  4.0 &  97.0 \\ 
 \WATET\                &  5.0 &   5.0 &  6.0 &  99.0 \\ 
  \WATP\                &  2.7 &   3.0 &  4.0 &   8.0 \\ 
                        &  2.7 &   1.2 &  4.0 &  11.5 \\ 
                        &  2.7 &   1.5 &  4.0 &  14.0 \\ 
                        &  2.7 &   3.0 &  2.0 &  64.5 \\ 
                        &  2.7 &   2.0 &  2.0 &  68.0 \\ 
                        &  2.7 &   1.0 &  2.0 &  78.0 \\ 
                        &  2.7 &   3.0 &  2.0 &  80.5 \\ 
                        &  5.0 &   6.0 &  4.0 &  96.0 \\ 
                        &  5.0 &  15.0 & 20.0 &  87.0 \\ 
   \WAT\       / W51N E1&  2.7 &   6.0 &  3.0 &   6.0 \\ 
                        &  2.7 &   1.0 &  5.0 &  11.0 \\ 
                        &  2.7 &   0.3 &  2.0 &  24.5 \\ 
                        &  2.7 &   3.5 &  1.0 &  45.0 \\ 
                        &  2.7 &   1.5 &  2.0 &  48.0 \\ 
                        &  5.0 & 400.0 &  5.0 &  57.5 \\ 
                        & 50.0 & 100.0 & 40.0 &  59.5 \\ 
                        &  2.7 & 250.0 &  7.0 &  64.0 \\ 
 \WATET\                &  5.0 &   3.0 &  5.0 &  57.5 \\ 
                        &  2.7 &   1.5 &  7.0 &  64.0 \\ 
  \WATP\                &  2.7 &   6.0 &  5.0 &   6.0 \\ 
                        &  2.7 &   2.0 &  5.0 &  11.0 \\ 
                        &  2.7 &   0.5 &  2.0 &  22.5 \\ 
                        &  2.7 &   0.3 &  1.0 &  45.0 \\ 
                        &  2.7 &   1.0 &  2.0 &  48.0 \\ 
                        &  5.0 &  17.0 & 20.0 &  50.0 \\ 
                        &  5.0 &   0.3 &  5.0 &  57.5 \\ 
                        &  2.7 &   4.0 &  4.0 &  64.0 \\ 